\documentclass[aps,prl,twocolumn,superscriptaddress]{revtex4-1}
\usepackage{graphicx}
\usepackage{multirow}
\renewcommand{\vec}[1]{\mathrm{\mathbf{#1}}}

\begin{document}
\title{The Shear Mode of Multi-Layer Graphene}
\author{P. H. Tan}
\author{W. P. Han}
\author{W. J. Zhao}
\author{Z. H. Wu}
\author{K. Chang}
\affiliation{Institute of Semiconductors, Chinese Academy of Sciences, Beijing 100083, China}
\author{H. Wang}
\author{Y. F. Wang}
\affiliation{Department of Physics, Nankai University, Tianjin 300071, China}
\author{N. Bonini}
\author{N. Marzari}
\affiliation{Department of Materials, University of Oxford,Oxford OX1 3PH, UK}
\author{G. Savini}
\author{A. Lombardo}
\author{A. C. Ferrari}
\affiliation{Engineering Department, Cambridge University, Cambridge CB3 OFA, UK}

\begin{abstract}
We uncover the interlayer shear mode of multi-layer graphene samples, ranging from bilayer-graphene (BLG) to bulk graphite, and show that the corresponding Raman peak measures the interlayer coupling. This peak scales from$\sim$43cm$^{-1}$ in bulk graphite to$\sim$31cm$^{-1}$ in BLG. Its low energy makes it a probe of near-Dirac point quasi-particles, with a Breit-Wigner-Fano lineshape due to resonance with electronic transitions. Similar shear modes are expected in all layered materials, providing a direct probe of interlayer interactions.
\end{abstract}
\maketitle

Single Layer Graphene (SLG) has high mobility and optical transparency, in addition to flexibility, robustness and environmental stability\cite{Geim07,bona}. As the knowledge of the basic properties of SLG increases, an ever growing effort is being devoted to a deeper understanding of Few Layer Graphene (FLG) samples\cite{bao,avouris,morpurgo}, and to their application in useful devices. For example, since SLG absorbs 2.3\% of the incident light\cite{nair}, FLG can be used to beat the transmittance of Indium Tin Oxide($\sim$90\%)\cite{bona}, and to engineer near-market transparent conductors\cite{bae}, exploiting the lower sheet resistance afforded by combining more than one SLG\cite{bona,bae}. Bilayer graphene (BLG) is a tunable band gap semiconductor \cite{castro}, tri-layer graphene (TLG) has a unique electronic structure consisting, in the simplest approximation, of massless SLG and massive BLG subbands\cite{jarillo,guinea,Kosh}. FLG with less than 10 layers do each show a distinctive band structure\cite{Kosh}. The layers can be stacked as in graphite, or have any orientation. This gives rise to a wealth of electronic properties, such as the appearance of a Dirac spectrum even in FLG\cite{latil}.

There is thus an increasing interest in the physics and applications of FLG. Optical microscopy can count the number of layers\cite{blake,cnano}, but does not offer the insights of Raman spectroscopy, being this sensitive to quasi-particle interactions\cite{Ferrari06}. Raman spectroscopy is one of the most useful and versatile tools to probe graphene samples\cite{Ferrari06,sscraman}. The measurement of the SLG, BLG, and FLG Raman spectra\cite{Ferrari06} triggered a huge effort to understand phonons, electron-phonon, magneto-phonon and electron-electron interactions, and the influence on the Raman process of number and orientation of layers, electric or magnetic fields, strain, doping, disorder, edges, and functional groups\cite{sscraman}.

The SLG phonon dispersions comprise three acoustic and three optical branches. A necessary, but not sufficient, condition for a phonon mode to be Raman active is to satisfy the Raman fundamental selection rule, i.e. to be at the Brillouin Zone centre, $\vec{\Gamma}$, with wavevector $\bf{q\approx0}$\cite{cardonabook}. SLG has six normal modes at $\vec{\Gamma}$: $A_{2u}+B_{2g}+E_{1u}+E_{2g}$\cite{nemanich0}. There are two degenerate in-plane optical modes, $E_{2g}$, and one out-of-plane optical mode $B_{2g}$\cite{nemanich0}. $E_{2g}$ modes are Raman active, while $B_{2g}$ is neither Raman nor IR active\cite{nemanich0}. In the case of graphite there are 4 atoms per unit cell, and only half of them have fourth neighbors that either lie directly above or below in adjacent layers. Therefore the two atoms of the unit cell in each layer are now inequivalent. This doubles the number of optical modes, and is responsible for the IR activity of graphite\cite{nemanich0}. All SLG optical modes become Davydov-doublets in graphite: $E_{2g}$ generates a IR active $E_{1u}$ and a Raman active $E_{2g}$, $B_{2g}$ goes into an IR-active $A_{2u}$, and an inactive $B_{2g}$. The zone boundary acoustic modes fold back to the zone centre as rigid layer modes: an optically inactive $B_{2g}$ and a Raman active $E_{2g}$. The acoustic modes remain $E_{2u}$ and $E_{1u}$\cite{nemanich0}. Thus for graphite\cite{nemanich0,mani} $\Gamma=2(A_{2u}+B_{2g}+E_{1u}+E_{2g})$. There are now two Raman active $E_{2g}$ modes, each doubly degenerate. The high frequency $E_{2g}$ mode is responsible for the well-known G peak, measured and discussed in thousands of papers to date for any carbon allotrope\cite{rsbook}.

Here we focus on the low energy $E_{2g}$ mode. This is a doubly degenerate rigid layer shear mode, involving the relative motion of atoms in adjacent planes. It was first measured in 1975 by Nemanich et al.\cite{nema1} in bulk graphite at$\sim$42cm$^{-1}$. We uncover the equivalent mode for FLGs, and show that it provides a direct measurement of the interlayer coupling. For this reason we name C the corresponding Raman peak. On one hand, the C peak energy, E(C)$\sim$5meV, is much lower than the notch and edge filter cuts of most Raman spectrometers, and its intensity is much smaller than the G peak. This explains why it was not seen thus far in FLG and, even for graphite, it was reported only in an handful of papers\cite{nema1,Hanftand89,Sinha90}, with no firm agreement on position and width. On the other hand, this makes it a probe of the quasi particles near the Dirac point.
\begin{figure}
\centerline{\includegraphics[width=95mm]{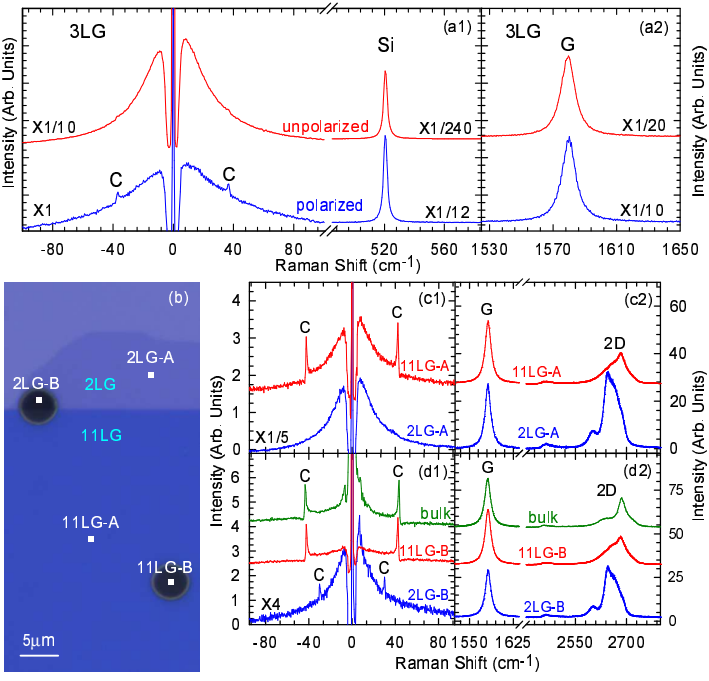}}
\caption{(a1) Unpolarized and polarized Raman spectra of 3LG on SiO$_2$/Si(001) in the C peak region and (a2) in the G peak region.(b) Optical micrograph of FLG sample. 2LG-A/11LG-A and 2LG-B/11LG-B denote supported and suspended flakes, respectively.(c1) S/AS spectra of supported samples in the C peak region.(c2) S-spectra of supported samples in the G/2D peaks region.(d1) S/AS spectra of suspended samples in the C peak region.(d2) S-spectra of suspended samples in the G/2D peak region}
\label{fig:1}
\end{figure}

Raman measurements are performed at 633nm in backscattering geometry using a Jobin-Yvon HR800 system. The laser plasma lines are removed using a bandpass filter, since those would appear in the same spectral range as the C peak. A typical laser power of 0.5mW is used to avoid sample heating. Detection of Raman modes down to$\sim$10cm$^{-1}$ is possible by using BragGrate notch filters with optical density 3, and with a spectral bandwidth$\sim$5-10cm$^{-1}$. Three filters for each excitation are necessary to suppress the Rayleigh signal, which is typically 10$^9$-10$^{12}$ stronger than the Raman signal\cite{cardonabook}. Ar gas is flown on the sample to remove the low-frequency Raman modes from the air. We use a 100$\times$ objective with NA=0.90. A 1800 lines/mm grating enables us to have each pixel of the charge-coupled detector cover$\sim$0.4cm$^{-1}$. A spectral resolution$\sim$0.5cm$^{-1}$ is estimated from the width of the Rayleigh peak.

The easiest way to get high quality SLG and FLG is by graphite exfoliation\cite{Novoselov04} on SiO$_2$/Si, to enhance visibility\cite{blake,cnano}. Often the Si is doped, to be used as back gate\cite{Novoselov04}. However, this poses a problem for low frequency Raman measurements. The incident light can excite carriers in doped Si, producing a strong background\cite{cardona}, that can overshadow the signal of FLG with less than 6 layers. One approach to overcome this issue is to perform polarized Raman measurements, since this background is strongly suppressed in cross polarization\cite{cardona}. Fig.\ref{fig:1}(a1) shows the unpolarized Raman spectrum (top graph) of 3LG on SiO$_2$/Si(001), as well as the polarized one (bottom graph) with incident light along [1 -1 0] and scattered light analyzed along [1 1 0]. A large substrate background is observed in the unpolarized measurement, while in the polarized one the Si mode and its low-frequency background are suppressed, thus revealing a peak$\sim$37cm$^{-1}$. However, polarized Raman yields low overall intensity, and affects the 2D to G ratio. Thus, in order to detect the C peak in unpolarized measurements, we take a different avenue. We use low doping Si (resistivity$\geq$2000$\Omega$.cm) and suspend the FLG on$\sim$2-5$\mu$m etched holes, as shown in Fig.\ref{fig:1}b. The number of layers is identified by a combination of 2D peak Raman spectroscopy\cite{Ferrari06} and optical contrast on the supported section of the flake\cite{blake,cnano}. Figs.\ref{fig:1}(c1,c2) plot the Stokes (S) and Anti-Stokes (AS) spectra for supported and suspended BLG, 11LG and bulk graphite. We use the notation NLG to indicate FLG with N layers. Thus 1LG=SLG; 2LG=BLG, 3LG=TLG, while, e.g., 11LG means 11 layers. In the suspended flakes the C peak is clearly seen. On the other hand, the supported ones show the Si background, Fig.\ref{fig:1}(c1). While for 11LG and bulk this does not overshadow the C peak, for fewer layers this covers the C peak, to the point that for supported BLG we cannot detect any C peak.
\begin{figure*}[tb]
\centerline{\includegraphics[width=160mm]{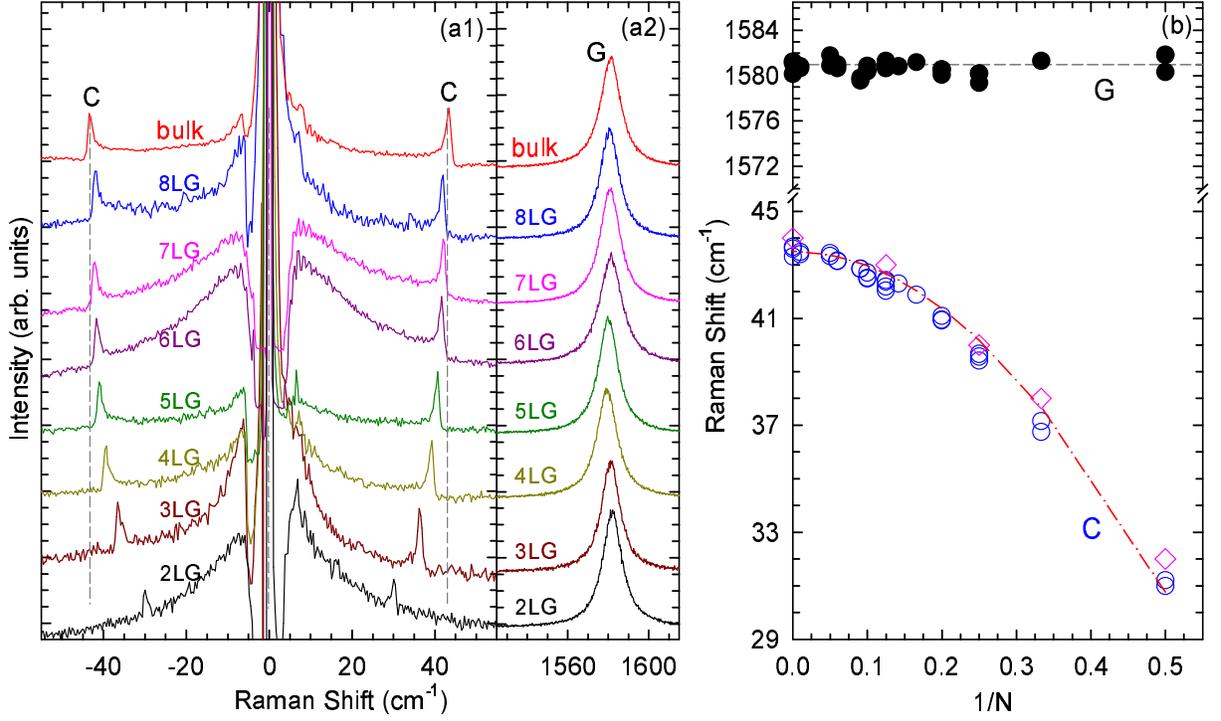}}\caption{(a1) S/AS C peak spectral region.(a2) G peak spectral region.(b) (solid black circles) Pos(G), (open blue circles) Pos(C), as a function of inverse layer number. The red dash-dotted line is a plot of Eq.\ref{eq:freq2}, open diamonds are DFT calculations}
\label{fig:2}
\end{figure*}

Given the low E(C), the S/AS intensity ratio is close to 1. Similar to the G peak\cite{Tan98}, but unlike D and 2D\cite{Tan98}, the S/AS C peaks are symmetric relative to the Rayleigh line. This allows us to precisely determine the C peak position, Pos(C), as [Pos(C)$_S$+Pos(C)$_{AS}$]/2. We get Pos(C)$\sim$31cm$^{-1}$ for BLG;$\sim$42.7cm$^{-1}$ for 11LG and$\sim$43.5cm$^{-1}$ in bulk graphite. By assuming a lorentzian lineshape, we derive the Full Width at Half Maximum, FWHM, to be FWHM(C)$\sim$1.2cm$^{-1}$. Considering the$\sim$0.5cm$^{-1}$ spectral broadening of our spectrometer, we derive an intrinsic linewidth$\sim$0.7cm$^{-1}$. From the S/AS ratio we estimate the local T on the sample as\cite{cardonabook}, $T=\hbar\omega/k_B ln\{I(C)_S/I(C)_{AS} \cdot \{[\omega_L + Pos(C)]/[\omega_L - Pos(C)]\}^4\}$, where $\hbar$ is the reduced Planck constant, $\omega_L$ is the laser frequency, $k_B$ is Boltzmann constant, $I(C)_S/I(C)_{AS}$ is the C peak S/AS intensity ratio. This gives T$\sim$300K, indicating negligible laser heating.

Figs.\ref{fig:2}(a1,a2) plot the Raman spectra for a set of samples of increasing thickness. Fig.\ref{fig:2}(b) shows the fitted Pos(G) and Pos(C) as a function of 1/N, where N is the number of layers. While Pos(G)$\sim$1581cm$^{-1}$ with no significant change with N, Pos(C) increases from BLG to bulk graphite. Note that in Fig.\ref{fig:2}(b) the spectral range used to plot the G and C peak data is the same (16.5cm$^{-1}$), thus the C peak shift with N is truly representative of its much stronger variation when compared to G. The ratio of peak heights at 633nm, I(C)/I(G) is$\sim$0.01, 0.06, 0.06 for BLG, 11LG and bulk graphite, while that of integrated peak areas, A(C)/A(G)$\sim$0.001, 0.006, 0.007. These change with excitation energy, e.g., A(C)/A(G)$\sim$0.0014 for bulk graphite at 532nm. Since these ratios depend on the Electron Phonon Coupling (EPC), this immediately indicates that EPC(C) is much smaller than EPC(G).

\begin{figure*}[tb]
\centerline{\includegraphics[width=140mm]{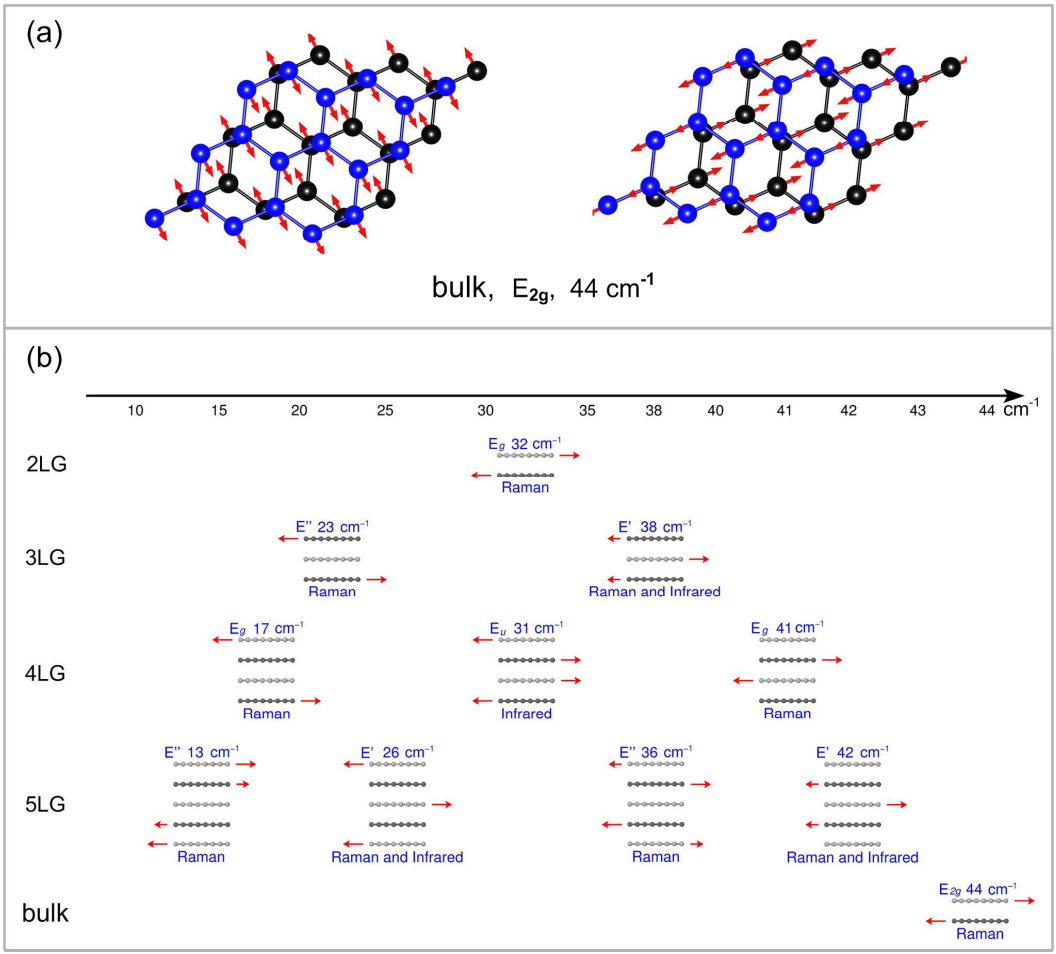}}
\caption{(a) The two degenerate E$_{2g}$ shear modes in graphite.(b) Symmetry and ab-initio frequencies for each shear mode. The Raman and IR active modes are identified.}
\label{fig:3}
\end{figure*}

The Pos(C) dependence on the number of layers can be explained considering a simple linear-chain model. For FLG with N layers, there are 2N atoms per unit cell. The corresponding in-plane optical modes consist of N degenerate pairs of in-plane stretching modes, and N-1 degenerate pairs of in-plane shear modes between neighboring layers. We assume that a layer interacts strongly only with adjacent layers and that the strength of this interlayer coupling is characterized by an inter-layer force constant per unit area, $\alpha$. The N-1 shear modes of a NLG can be computed by diagonalizing the corresponding $N\times N$ (tridiagonal) dynamical matrix. The frequency $\omega_i$ (in Hz) of the $i$-th vibrational mode is given by:
\begin{equation}
\omega^2_i=2\frac{\alpha}{\mu}\Big\{1-\cos\Big[\frac{(i-1)\pi}{N}\Big]\Big\}
\label{eq:freq}
\end{equation}
\noindent where i=2,...N. $\mu$=7.6$\times$10$^{-27}$Kg/{\AA}$^2$ is the SLG mass per unit area. The corresponding $i$-th displacement eigenvector $v^{(i)}_j$ is given by:
\begin{equation}
v^{(i)}_j = \cos\Big[\frac{(i-1)(2j-1)\pi}{2N}  \Big]
\label{eq:displ}
\end{equation}
\noindent where $j$ labels the layers. The highest frequency mode (for i=N) is Raman-active. Here adjacent layers move out-of-phase in the direction parallel to the planes. In the case of graphite $N \to \infty$ and $\omega_{\infty}=Pos(C)_{\infty} \times (2 \pi c)=2 \sqrt {\alpha / \mu}$, where c is the speed of light in cm/s. This is the doubly degenerate E$_{2g}$ shear mode responsible for the C peak, see Fig.\ref{fig:3}(a). Thus, Pos(C)$_N$ (in cm$^{-1}$) for a NLG is given by Eq.\ref{eq:freq} setting i=N:
\begin{equation}
Pos(C)_N=\sqrt{2}\sqrt{\frac{\alpha}{\mu}}\sqrt{1+\cos\Big(\frac{\pi}{N}\Big)}\times \frac{1}{2 \pi c}
\label{eq:freq2}
\end{equation}
In BLG, N=2, and $Pos(C)_2=\sqrt{2} \sqrt {\alpha / \mu}\times \frac{1}{2 \pi c}$, i.e. $\sqrt{2}$ smaller than Pos(C)$_{\infty}$, corresponding to bulk graphite, in excellent agreement with the experiments. In fact, the dash-dotted line in Fig.\ref{fig:2}(b) shows that Eq.\ref{eq:freq2} describes all the experimental data, thus validating our simple model. The only unknown parameter in Eq.\ref{eq:freq2} is the interlayer coupling strength. By fitting the experimental data we can directly measure it. We get $\alpha\sim$12.8$\times$10$^{18}$N/m$^3$. Thus, in Bernal stacked FLG, the C mode hardening is not due to a variation of interlayer coupling, but rather to an increase of the overall restoring force going from BLG to bulk graphite. For a given N, we expect variations of Pos(C) if the interlayer coupling is modified, e.g. by changing the spacing or relative layer orientation (in the latter case we also expect mode splitting).

These results are further confirmed by density functional theory (DFT) calculations. These are performed using DFT and density-functional perturbation (DFPT) theory as implemented in the PWSCF package of the QUANTUM-ESPRESSO distribution, within the Local Density Approximation(LDA), and ultrasoft pseudopotentials generated using the RRKJ approach. The cutoffs are 40Ry for the wave functions, and 480Ry for the charge density. The Brillouin zone is sampled on a $42\times42\times16$ Monkhorst-Pack mesh for bulk graphite and $42\times42\times1$ for SLG and BLG. NLG are modeled using supercell configurations, with periodic replicas separated by 10$\textrm{\AA}$ vacuum in the perpendicular direction. The electron-phonon and phonon-phonon matrix elements, as well as the anharmonic contribution to the C mode linewidth, are computed using the approach of Ref.\cite{bonini}. The EPC contribution to the linewidth is computed using an interpolation based on maximally-localized Wannier functions as implemented in the EPW code\cite{epw_ref}. This is a computationally efficient approach allowing very fine sampling of the Brillouin zone (meshes of several million points are needed to get accurate phonon linewidths).

Fig.\ref{fig:3}(b) plots the in-plane shear modes for 2-5LG and bulk graphite. For a given N, there are N-1 shear modes, either Raman or IR active or both, but, for N$>$2, N-2 of those have a different displacement pattern compared to C, since not all the neighboring layers vibrate out-of-phase. The highest frequency Raman active mode corresponds to the C peak. We expect the other Raman active modes to have a much weaker intensity compared to the C peak, as a result of a smaller EPC, also confirmed by DFT. More work is needed to detect those modes.

We get an excellent agreement between our DFT frequencies and the experimental data, as indicated by the open symbols in Fig.2(b). This might seem surprising, since local or semi-local exchange correlation functionals may not properly describe Van Der Waals interactions\cite{Toulouse04}. However, it was shown that in bulk graphite all phonon dispersions are well described by DFT, both in LDA and in the generalized gradient approximation (GGA), even in absence of Van der Waals (VdW) interactions in the functional, provided that the correct geometry (i.e. interlayer spacing) is used\cite{mounet}. This occurs because VdW interactions can give a significant contribution to the total energy (hence determining the c/a ratio, where c is the interlayer spacing and a the in plane lattice constant), but give a negligible contribution to second derivatives (i.e. the phonons). Hence, if the correct geometry is used, phonon dispersions are well reproduced. LDA does provide excellent geometries: c/a for graphite is 2.725, 2.74, 2.91, for experiments, LDA, VdW-DFT, respectively. For BLG, LDA and VdW-DFT give c/a=2.74, 2.90, i.e. VdW-DFT predicts the same spacing in bulk graphite and BLG, while LDA consistently predicts a smaller value than VdW-DFT, but in excellent agreement with experiments. This is confirmed, independently, for bulk graphite and any FLG, by the very good agreement between our LDA calculations and the measured FLG C modes.

\begin{figure}[tb]
\centerline{\includegraphics[width=90mm,clip]{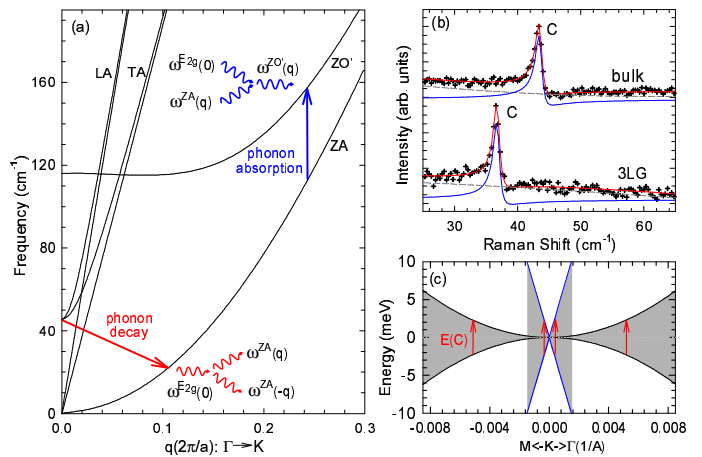}}
\caption{(a)Schematic representation of the anharmonic decay channels for the C mode in bulk graphite.(b) C peak for 3LG and bulk graphite, with BWF fit.(c) Schematic band structure of 3LG close to K. The gray regions highlight transitions near K that could resonate with the C mode. Red arrows indicate transitions with the same energy as the C mode, E(C).}
\label{fig:4}
\end{figure}

We now consider FWHM(C). Two factors contribute to the linewidth of the E$_{2g}$ Raman modes in graphene and graphite: the EPC term\cite{lazzeri,pisana} and anharmonic phonon-phonon interactions\cite{bonini}. In the absence of doping, FWHM(G)$\sim$12-14cm$^{-1}$ in SLG and bulk graphite, mostly due to the dominant EPC contribution\cite{lazzeri,pisana}, the phonon-phonon one being$\sim$1.7cm$^{-1}$\cite{bonini}. The experimental FWHM(C) is much smaller not just compared to the overall FWHM(G), but also with respect to the non-EPC component of FWHM(G). This immediately indicates a much smaller EPC(C) than EPC(G), consistent with the much smaller C peak intensity. Our DFT calculations give FWHM(C)$\sim$0.3cm$^{-1}$ at 300K in graphite, in reasonable agreement with experiments. The EPC contribution being$\sim$0.05cm$^{-1}$, and the phonon-phonon one$\sim$0.25cm$^{-1}$. The anharmonic term consists of three-phonon decay (30$\%$ of the total anharmonic linewidth at 300K) and absorption (70\%) processes. The shear mode splits mainly into two out-of-plane ZA bending modes, at {\bf q} and -{\bf q}, close to $\vec{\Gamma}$. The absorption processes are dominated by the merging of the shear mode and a ZA mode into an out-of-plane ZO' bending mode (the prime indicates an optical mode where the two atoms in each layer of the unit cell of graphite vibrate together, but out-of-phase with respect to the two atoms of the other layer), see Fig.\ref{fig:4}a. We expect the anharmonic linewidth not to change significantly in NLG, since the available phase space of decay/absorption channels in these systems is very similar to graphite. Also, our calculations for BLG, TLG and 4LG show that the EPC contribution to the linewidth is nearly independent of the number of layers. Thus, DFT indicates the overall FWHM not to change significantly with N, in agreement with experiments. Note that, if we take a 4LG as an example, the EPC contribution to FWHM(G) is$\sim$150 times bigger than for FWHM(C). In turn, this is$\sim$15 times bigger than the EPC contribution to the other Raman active shear mode$\sim$17cm$^{-1}$, confirming the expectation that the other C modes would be challenging to detect.

We now examine more closely the C peak shape. Fig.\ref{fig:4}(b) shows that it can be well fitted with a Breit-Wagner-Fano (BWF). In general, this arises as quantum interference between a Raman allowed phonon and a continuum of Raman active electronic (or multiphonon) transitions\cite{klein}. The BWF lineshape is\cite{klein}:
\begin{equation}
I(\omega)=I_0
\frac{[1+2(\omega-\omega_0)/(q\Gamma)]^2}{[1+4(\omega-\omega_0)^2/\Gamma^2]}
\label{eq1}
\end{equation}
\noindent where I$_0$, $\omega_0$, $\Gamma$ and 1/$|q|$ are the intensity, uncoupled mode frequency, broadening parameter and coupling coefficient. The peak maximum is at $\omega_{max}$=$\omega_0+\Gamma/2q$, while its FWHM=$\Gamma(q^2+1)/|q^2-1|$. In the limit 1/$q$$\rightarrow$0, a Lorentzian lineshape is recovered, with FWHM=$\Gamma$ and $\omega_{max}$=$\omega_0$. Pos(C) in the BWF fit is $\Gamma/2|q|$ ($\sim$0.3cm$^{-1}$) higher than in a Lorenzian fit.

We find a smaller 1/$|q|$ when we use a laser power high enough to shift the G peak, i.e. to heat the sample. Thus, in our low power experiments, the possible laser-induced electron-hole plasma is not the cause of the observed BWF lineshape. We also find that the C mode of bulk graphite at 77K has the same q as at room temperature, in contrast to what expected if the BWF would be due to a multiphonon resonance\cite{Dresselhaus82}. We thus attribute the BWF lineshape to quantum interference between the C mode and a continuum of electronic transitions near the K point. The band structure of Bernal-stacked FLGs can be decomposed into groups of BLG bands, with different effective masses, plus- for odd layer numbers- a pair of SLG bands\cite{Kosh}. Fig.\ref{fig:4}(c) plots, as an example, the schematic band structure of 3LG in a range of the order of the phonon energy of the C mode, E(C), and identifies electronic transitions that can couple with the C mode. Because the density of states with energy higher than E(C) is much larger than that with energy smaller than E(C), $q$ is not expected to change significantly from BLG to bulk graphite, in agreement with our findings. If the Fermi energy, E$_F$, is larger than E(C)/2, the resonance with the C mode would become weaker, eventually leading to the disappearance of the BWF profile.

As discussed above, FWHM(G) is much larger than FWHM(C), due to the much larger EPC and phonon-phonon contributions. The EPC dominates FWHM(G), and the G peak is always lorentzian.

In summary, we uncovered the Raman signature of the interlayer shear mode of FLG. Graphite is not the only layered material. Transition metal dichalcogenides, transition metal oxides, and other compounds such as BN, Bi$_2$Te$_3$, and Bi$_2$Se$_3$ can also be exfoliated to produce a whole range of two-dimensional crystals, that are just beginning to be investigated\cite{coleman}. Similar shear modes are expected in all these materials, and their detection will provide a direct probe of interlayer interactions.

\textbf{Acknowledgments}
We thank Ed McCann, Mikito Koshino, and  Timo Thonhauser for useful discussions. This work was supported by NSFC grants 10874177, 10934007, 60878025, special funds for the Major State Basic Research of China 2009CB929301, the doctoral programme of higher education of China 20100031110004, the ERC grant NANOPOTS, EPSRC grant EP/G042357/1, a Royal Society Wolfson Research Merit Award, EU grants RODIN and Marie Curie ITN-GENIUS (PITN- GA-2010-264694), and Nokia Research Centre, Cambridge.

\end{document}